\documentstyle[psfig,sprocl]{article}

\def\be{\begin{equation}}
\def\ee{\end{equation}}
\def\bea{\begin{eqnarray}}
\def\eea{\end{eqnarray}}
\begin{document}

\title{\large A Deep Probing of the Coma cluster}

\author{C. ADAMI \& A. MAZURE}

\address{IGRAP: LAS, T. du Siphon, 
13012 Marseille, France, E-mail:adami@astrsp-mrs.fr} 

\author{F. DURRET \& C. LOBO}

\address{IAP, 98bis Bd Arago, 
F 75014, France, E-mail: durret@iap.fr} 

\author{B. HOLDEN}

\address{University of Chicago, Dept of Astronomy and Astrophysics, 
5640 S. Ellis Av., USA} 

\author{R. NICHOL}

\address{CMU, Dept of physics, Wean Hall 5000, 
Forbes Avenue, PA 15213-3890, USA} 


\maketitle\abstracts{We present here results from a deep 
spectroscopic survey of the
Coma cluster of galaxies (29 galaxies between 18.98 $\le$ $m_R$ $\le$ 21.5). Only
1 of these galaxies is within Coma compared to an expected 
6.7 galaxies computed from nearby control fields.
This discrepancy potentially indicates that Coma's faint end 
luminosity function has been
grossly overestimated and raises concerns about the validity of using 2D
statistical subtraction to correct for the background galaxy population when 
constructing cluster luminosity functions.}

\section{Introduction}\label{subsec:intro}

The Coma Cluster is one of the most studied clusters of galaxies in
the sky. Its
high richness and low redshift have led it to become the
archetypal target for both observational and theoretical cluster studies. One
aspect, however, of Coma that has yet to be fully investigated
is the distribution of faint cluster galaxies. Using
CFHT and the MOS spectrograph, we have partially re-adressed this
shortfall by obtaining spectroscopy for a sample of
18.98 $\le$ $m_R$ $\le$ 21.5
galaxies in the direction of the core of Coma. We selected 105 targets
within this magnitude range from the deep field already imaged by Bernstein
et al. (1995: B95).

\section{Scentific Goals}\label{subsec:goals}

There are 3 scientific goals of this work: {\it 1)}
We have found 3 faint galaxy overdensities 
in the deep optical imaging
data of B95 that may
be coincident with irregularities in Coma's X-ray
emission (White et al. 1993). These small scale 
substructures need to be confirmed using
redshift information; {\it 2)} We do not fully know the contribution 
of galaxies
fainter than the photographic plate limit (Godwin et al. 1983) to the
overall light of Coma and whether this contribution can
increase the known baryonic mass in the
cluster; {\it 3)} Finally, the exact shape of Coma's luminosity function
(B95, Lobo et al. 1996) remains uncertain, varying
from a single atypical Schechter function with a steep faint-end slope,
to a Gaussian at bright magnitudes and a power-law at the
faint-end. These uncertainties come from the reliability of the 2D
statistical correction for the background, which is prone to severe
bias. Coma is an excellent first candidate to check these procedures.

\section{Results}
%\subsection{General observational results}\label{subsec:gene}

In March 1997, we 
spectroscopically observed only 29 of the 105 target galaxies discussed above
(bad weather).
In addition, we successfully
observed 5 other faint galaxies outside the B95 area.
The apparent magnitudes of the observed sub-sample 
are uniformly distributed among the 105 available objects. 
All observed objects were spectroscopically confirmed to be
galaxies {\it i.e.} we did not discover an misclassified
stars or globular clusters. 
Our success rate in obtaining redshifts from the spectra was 70\%
with the typical error on each redshift of $\delta z= 0.001$. The 
majority of the
redshifts are uniformly distributed along the line of sight
between $0.17<z<0.88$ (fig.~1); significantly beyond the
redshift of Coma (Biviano et al. 1996). We found only 1
galaxy in Coma (z=0.02).
We also detected a significant peak of 6 galaxies at
z$\simeq0.5$ ($\delta z=0.02$). The probability of having 6
galaxies clustered in redshift  -- 
 compared to a  uniformly
distributed -- is $<2\times10^{-6}$ \%.  

For the 29 observed galaxies 
in the B95 area, we would have predicted 6.7 should have been part of the
velocity field of the cluster. The large error bar on
this prediction does not prohibit a conclusive statement about
the errors in the 2D statistical background
subtraction method. However, it is very important to check this result with
more redshifts since the 2D statistical subtraction
of background galaxies is now standard practice 
for constructing cluster luminosity functions; both distant and nearby.

\begin{figure}
%\begin{center}
\vbox
{\psfig{figure=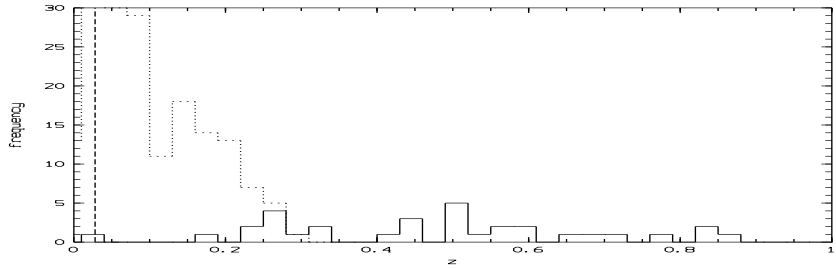,height=4.0cm,width=12.0cm,angle=270}}
\caption{Redshifts known (dashed), the 34 new ones (solid
) and Coma (long-dashed)}
\label{}
%\end{center}
\end{figure}

%\subsection{A physical structure at z=0.5 beyond Coma ?}\label{subsec:amas}

One of the sub-clumps mentioned in section 2 is coincident with the 
redshift
peak at z$\simeq$0.5 identified above. 
The spatial extension of this structure is 110 kpc
(H$_0$=100 \& q$_0$=0) and the probability of having 6 galaxies
clustered in both redshift and angle now becomes $<2\times10^{-9}$ \% (if we
assume a random distribution in redshift and angle).  The mean velocity of
this structure is 153790 km.s$^{-1}$ with a dispersion calculated at
z=0 of 1315 km.s$^{-1}$. The 6 spectra all
have observed absorption lines which may be a hint of an 
interaction of these galaxies with a dense medium. These observations are 
consistent with a group, or cluster 
of galaxies, beyond Coma at z=0.5. This possible
system does not correspond to any group from Hickson
(1982) or Barton et al. (1996).

\end{document}